\documentclass{ws-ijmpd}
\usepackage[super,compress]{cite}
\begin{document}


%
\catchline{}{}{}{}{}
%

\title{COSMIC ACCELERATION FROM VARYING MASSES IN FIVE DIMENSIONS}

\author{P.H.R.S. MORAES}

\address{ITA - Instituto Tecnol\'ogico de Aeron\'autica - Departamento de F\'isica, 12228-900, S\~ao Jos\'e dos Campos, S\~ao Paulo, Brazil\\
moraes.phrs@gmail.com}



\maketitle

\begin{history}
\end{history}

\begin{abstract}
Much effort has been made in trying to solve or at least evade the inconsistencies that emerge from general relativity as the framework for a cosmological model. The extradimensional models rise as superb possibilities on this regard. In this work I present cosmological solutions for Wesson's Space-Time-Matter theory of gravity. A relation between mass variation at cosmological scales and the expansion velocity of the universe is obtained. Such a relation yields novel features on Space-Time-Matter theory of gravity, which are carefully discussed.
\end{abstract}



\tableofcontents

\section{Introduction}\label{sec:int}

Extradimensional gravitational models have been proposed as General Relativity (GR) alternative for a long time. They are reported in the literature with different forms of describing our world as a five-dimensional (5D) rather than four-dimensional (4D) space-time\cite{clifton/2012}. Among the most important of them, one can quote Kaluza-Klein (KK) gravity\cite{overduin/1997}, which considers the universe as empty in 5D, with the 4D matter content arising purely as a geometrical manifestation. KK field equations (FEs) contain both the 4D Einstein's GR FEs and Maxwell's electromagnetism equations in the absence of a source; therefore KK theory is said to unify gravitation and electromagnetism. In order to unify also the other forces of nature - namely the weak and strong nuclear forces - in a single theory, it is common to consider more extra dimensions, contemplating the so-called string theories\cite{maldacena/1999,maharana/2013,sen/2005,berman/2014}.

The description of our world in 5D, as in KK or braneworld models\cite{randall/1999, randall/1999b, sahni/2003, afonso/2007, dvali/2001}, may shed some light in standard cosmology shortcomings, as cosmological constant (CC), coincidence, dark matter and hierarchy problems\cite{clifton/2012,rubakov/1983,dvali/2003,vinet/2004,lue/2003,bogdanos/2007}. 

In this paper, I am going to work with the extradimensional model named Space-Time-Matter theory of gravity (STM) \cite{wesson/1983,wesson/1984, wesson/1985, wesson/1990, wesson/1992, wesson/2015}. The basis of STM consists in taking the fundamental constants $c$ and $G$ - the speed of light and the Newtonian gravitational constant, respectively - and combine them with the rest mass $m$ of objects, yielding a length $Gm/c^{2}$ in the line element. Note that in the same way that in 4D the time $t$ is converted to a coordinate in order to form space-time, in 5D STM, $m$ is converted to a coordinate $Gm/c^{2}$, giving rise to space-time-matter. STM proposes, then, the rest mass of objects, $m$, to slowly vary with time at a rate governed by the age of the universe\footnote{In fact, when some quantities at first sight considered as constants appear to evolve with time, it is natural to assume that their cosmic drift rate should be proportional to the Hubble time\cite{fritzsch/2015}.}. Indeed, in such an approach, the extra dimension is a representation of matter.

It is common to find in the literature endorsements for time-varying masses at a fundamental level\cite{langenfeld/2009,saito/1997,narison/2010,king/2008,shelkovnikov/2008}. Recently, a limit on the variation of the light quark mass has been obtained, as well as its consequences on nuclear force and big-bang nucleosynthesis\cite{berengut/2013}. Moreover, at large scales, it has been studied the variation of the so-called initial mass function of the universe\cite{kroupa/2001}. What has been proposed by P.S. Wesson through STM is that those variations would imply an increase in the measure of separation between two arbitrary events. Such an increase becomes more relevant at cosmological scales.

It is also common to find in the literature endorsements for time-varying ``constants", such as the fine structure constant\cite{murphy/2003,rahmani/2014}, or more specifically, the speed of light\cite{barrow/1999,barrow/1999b,moraes/2015b} or the newtonian gravitational constant\cite{barrow/2002,tripathy/2013}. Indeed, the variation of masses is not entirely independent of the variation of ``constants". If the fine structure constant $\alpha$ is, in fact, a function which slowly varies with time, the masses of all nucleons are allowed to vary as well, since the interaction responsible for the variation of $\alpha$ should couple radiatively to nucleons\cite{fritzsch/2015}. As a result, we expect the masses of protons and neutrons to vary with time\cite{olive/2002}. It should be noted that, recently, a lot of efforts have been made in approaching the variation of fundamental ``constants" with time as well as searching for observational evidences of it\cite{calmet/2015,barrow/2015,ward/2015,basilakos/2015,capozziello/2015}.

The variation of masses was explicitly studied in STM as a consequence of the creation of neutral fundamental particles\cite{israelit/2007,israelit/2008}. Recently, the subject of the continuous matter creation in the universe - which, naturally, would cause the mass variation effect - has been considered within the perspective of an accelerating universe\cite{lima/2010,lima/2011} and thermodynamics\cite{harko/2014}, the latter, been a consequence of non-conservative energy-momentum tensor theories, as the $f(R,T)$ theories of gravitation\cite{harko/2011}, for instance. 

A more extensive description of STM will be given in Section \ref{sec:stm}. For now, it is worth mentioning some applications that have been done to it so far. Without making any assumptions about the physical content of the universe, it has been obtained an equation of state (EoS) which describes the radiation-dominated era of the universe\cite{fukui/1992}. Static and anisotropic cosmological solutions were also obtained\cite{bermam/1995,roque/1991}. The age of the universe was calculated for homogeneous and isotropic models and it was shown that such a value does not depend on the nature of its matter content\cite{lima/1994}. General solutions for a homogeneous and isotropic universe were obtained, and some of those show compactification of the fifth dimension in different limits\cite{deleon/1988}.

Note that, recently, STM contributions and extensions have also been presented. For instance, a global differential geometric structure for STM was formulated, in which the model was extended to an arbitrary number of dimensions\cite{betounes/2013}. The fully general equations of motion in a covariant form were presented\cite{bejancu/2013}; and based on it, the FEs in a general KK space with electromagnetic potentials have been obtained\cite{bejancu/2013b}. Moreover, an expression for the rest mass of test particles and its variation along the 4D part of the 5D geodesic was provided\cite{deleon/2003}, while the question of how an observer in 4D perceives the 5D geodesic motion in STM has been investigated\cite{deleon/2004}. A connection of STM with dark matter and particle dynamics has been discussed\cite{liko/2004}. Other recent applications of STM can also be found\cite{wesson/2013, israelit/2009, overduin/2007, deleon/2007, xu/2006, deleon/2006, wesson/2006}.
 
The scope of this work is to obtain STM cosmological solutions from a general 5D metric. I also present and discuss some novel features on such a theory of gravity, as well as their successful cosmological consequences. The paper is organized as follows. In Section \ref{sec:stm} I highlight some relevant fundamentals of STM. In Section \ref{sec:fes} I present the FEs of the model and their solutions. In Section \ref{sec:novel} I derive new features on STM and present their physical consequences. In Section \ref{sec:dis} I discuss the results. 
 
\section{The Space-Time-Matter theory of gravity}\label{sec:stm}	

The STM considers the universe as a 5D manifold, with the fifth coordinate being the mass as $Gm/c^{2}$\cite{wesson/1983,wesson/1984, wesson/1985, wesson/1990, wesson/1992, wesson/2015}. As another variable-gravity theories\footnote{Variable-gravity theories share the property that the strength of the gravitational force varies with time.}, STM proposes the rest mass of objects, $m$, to slowly vary with time at a rate governed by the age of the universe. STM was constructed in such a way that if the quantity $(G/c^{3})dm/dt$ is null, masses are constants and the Einstein's theory of GR is recovered. Note that while astrophysical systems of small size (as stars) appear to have scales when observed over time scales typical of human research, there is no evidence that these scales remain fixed over time intervals comparable to the Hubble time.

P.S. Wesson, the STM founder, had pointed two grounds for suggesting that the length $Gm/c^{2}$ might be considered a coordinate: (1) since his desire was to create a variable-gravity theory, a variable mass was coherent and plausible; (2) just as the existence of $c$, a fundamental constant of GR, suggests a coordinate to be defined as $ct$, the existence of a second fundamental constant on GR, namely $G$, might suggest $Gm/c^{2}$ to be defined as a coordinate.

Wesson verified that the addition of such a fifth dimension to the usual four does not alter noticeably the geometry of the universe at non-cosmological scales. However, in cosmology, the fifth dimension can have relevant consequences, since in a homogeneous universe, the mass increases in proportion to the distance $r$ as $m\propto r^{3}$. Therefore, in the realm of cosmology, STM differs noticeably from the standard 4D theory of gravitation.

Departing from KK theory mentioned above, the fifth dimension in STM was not introduced in order to incorporate electromagnetism into relativity, but to incorporate mass into relativity in a way consistent with scale invariance.  

The FEs of the model might assume two distinct forms. They depend on how one relates the geometry considered to the properties of matter. One form can be taken simply by the 5D version of the Einstein's GR 4D ones, i.e. $G_{AB}=(8\pi G/c^{4}) T_{AB}$, with $A,B$ running from $0$ to $4$. However, it has been shown\cite{wesson/1992b}, as an alternative, that a 5D theory does not necessarily need an explicit energy-momentum tensor; the extra terms - due to the extra dimension - of the 5D Einstein's tensor may work as an induced energy-momentum tensor. Examples of the latter case have been recently developed \cite{moraes/2014,moraes/2015}. In this work, I am going to solve the FEs of the former case.

There is a special case for the STM metric, named canonical metric\cite{mashhoon/1994}, whose form yields exactly the same dynamics as GR, ensuring, therefore, that the 4D theory agrees with the classical solar system tests. Departures from such a metric normally yields a dependence of the cosmological ``constant" on the fifth coordinate. Such a dependence can solve the 4D CC problem\cite{overduin/2007,mashhoon/2004,wesson/2013b} mentioned above.

Furthermore the geodesic equation derived for STM yields exactly those of GR plus a new one, restricted to the 5D theory, which is

\begin{equation}\label{eqn:stm1}
\frac{d^{2}m}{dt^{2}}=0.
\end{equation}
Note that (\ref{eqn:stm1}) implies that the rate at which mass is varying is constant, which is in accordance, for instance, with a model-free method used to report a stable limit on the fractional temporal variation of the proton to electron mass ratio\cite{shelkovnikov/2008} and stringent constant values found for the variation of the light quark mass\cite{berengut/2013}.

At this time, the reader should be aware of the fact that the variation of masses at cosmological scales, which in STM is in the same footage of the other space-time dimensions, arise as a consequence of mass variations at a subatomic level, as one would expect.

Summarizing, in STM, $(G/c^{3})dm/dt\neq0$ characterizes the extra dimension by describing the rate of change of mass with time. By allowing the rest mass to vary, STM get rid of the fundamental mass scale of GR. By taking $dm/dt=0$, one should automatically recover GR. Moreover, note that providing $dm/dt\ll1$, STM agrees with observations\cite{wesson/1983}.

\section{The Space-Time-Matter field equations and their cosmological solutions}\label{sec:fes}

The STM FEs will be constructed by considering GR FEs in 5D. They will be given by $G_{AB}=8\pi T_{AB}$, for which the Einstein tensor is $G_{AB}=R_{AB}-(1/2)Rg_{AB}$, $R_{AB}$ is the Ricci tensor, $R$ is the Ricci scalar, $g_{AB}$ is the 5D metric, $T_{AB}=diag(\rho,-p,-p,-p,0)$ is the energy-momentum tensor of a perfect fluid, in which $\rho$ and $p$ are the density and pressure of the universe, respectively, and $A,B$ run from $0$ to $4$. I will assume since Wesson's theory is essentially a parametrisation of rest mass, one must have the fifth term on $T_{AB}$ equal to zero\cite{khadekar/2007,moraes/2014,moraes/2015,billyard/2001}. Moreover, throughout this work I am going to assume units such that $c=G=1$. 

The model FEs will be developed from the following metric

\begin{equation}\label{eqn:fes1}
dS^{2}=\beta^{2}(t,m)dt^{2}-\sigma^{2}(t,m)(dx^{2}+dy^{2}+dz^{2})-\varsigma^{2}(t,m)dm^{2},
\end{equation}
in which I have assumed the signature originally proposed by Wesson. Such a generic form for the 5D metric has been broadly used in STM\cite{moraes/2014,moraes/2015,deleon/2001,liu/2001,chatterjee/1987,fukui/1992}. It benefits from the fact that the scale factors depend on both time and extra coordinate, departing from the usual time(only)-dependence found in KK models\cite{bermam/1995,pradhan/2008,sharif/2011}. 

Let me assume, from now on, $\beta(t,m)=1$. Such an assumption represents merely a rescaling of time coordinate and has already been considered in Wesson's theory \cite{moraes/2015,chatterjee/1987,halpern/2001}. Moreover, it is usual to assume that the scalar of expansion of a 5D metric is proportional to its shear scalar in such a way one can write $\varsigma (t,m)=\sigma (t,m)^{n}$, with $n\neq 1$ being a constant\cite{moraes/2014,collins/1980,borkar/2015,rao/2015,rao/2015b,rao/2015c,reddy/2014,reddy/2013,reddy/2013b}. By doing so, the diagonal terms of the FEs are

\begin{equation}\label{eqn:fes2}
\left[(n+1)\left(\frac{\dot{\sigma}}{\sigma}\right)^{2}-\frac{\sigma^{\dagger\dagger}}{\sigma^{2n+1}}+(n-1)\left(\frac{\sigma^{\dagger}}{\sigma^{n+1}}\right)^{2}\right]=\frac{8}{3}\pi\rho,
\end{equation}
\begin{equation}\label{eqn:fes3}
(n+2)\frac{\ddot{\sigma}}{\sigma}+(n^{2}+n+1)\left(\frac{\dot{\sigma}}{\sigma}\right)^{2}-2\frac{\sigma^{\dagger\dagger}}{\sigma^{2n+1}}+(2n-1)\left(\frac{\sigma^{\dagger}}{\sigma^{n+1}}\right)^{2}=-8\pi p,
\end{equation}
\begin{equation}\label{eqn:fes4}
\frac{\ddot{\sigma}}{\sigma}+\left(\frac{\dot{\sigma}}{\sigma}\right)^{2}-\left(\frac{\sigma^{\dagger}}{\sigma^{n+1}}\right)^{2}=0.
\end{equation}
In Eqs.(\ref{eqn:fes2})-(\ref{eqn:fes4}), dots and daggers stand for partial derivatives with respect to time and mass, respectively. By using (\ref{eqn:fes4}), it is straightforward to show that the model FEs reduce to

\begin{equation}\label{eqn:fes5}
3\left[(n-1)\frac{\ddot{\sigma}}{\sigma}+2n\left(\frac{\dot{\sigma}}{\sigma}\right)^{2}-\frac{\sigma^{\dagger\dagger}}{\sigma^{2n+1}}\right]=8\pi\rho,
\end{equation}
\begin{equation}\label{eqn:fes6}
(3n+1)\frac{\ddot{\sigma}}{\sigma}+(n^{2}+3n)\left(\frac{\dot{\sigma}}{\sigma}\right)^{2}-2\frac{\sigma^{\dagger\dagger}}{\sigma^{2n+1}}=-8\pi\omega\rho,
\end{equation}
for which the EoS $p=\omega\rho$, has been adopted, with $\omega$ being a constant.

A universe dominated by matter has $\omega=0$. By assuming such an EoS in (\ref{eqn:fes6}), one obtains as a solution for $\sigma(t,m)$ the following:

\begin{equation}\label{eqn:cs1}
\sigma=\{[(n+3)^{2}-8]t\}^{\eta}m,
\end{equation}  
with $\eta(n)=(3n+1)/[(n+3)^{2}-8]$. By substituting (\ref{eqn:cs1}) in (\ref{eqn:fes5}), one obtains

\begin{equation}\label{eqn:cs2}
\rho=\frac{3}{8\pi}[2\eta^{2}n+\eta(\eta-1)(n-1)]t^{-2}.
\end{equation}
Note that although working in the 5D STM with the metric potentials depending on both $t$ and $m$, solution (\ref{eqn:cs2}) does not depend on the fifth coordinate, being restricted, then, to the usual 4D space-time.

From (\ref{eqn:cs1}), the mean scale factor $a(t,m)=[\sigma^{3}(t,m)\varsigma(t,m)]^{1/4}$ derived for the model is

\begin{equation}\label{eqn:cs3}
a(t,m)=\{[(n+3)^{2}-8]tm\}^{\eta\left(\frac{n+3}{4}\right)},
\end{equation}
from which one obtains

\begin{equation}\label{eqn:cs4}
H=\eta\left(\frac{n+3}{4}\right)t^{-1}
\end{equation}
for the Hubble parameter. 

So far, from Eqs.(\ref{eqn:cs2}) and (\ref{eqn:cs4}), STM has presented well behaved cosmological features. Similar features have already been found in others Wesson's theory applications\cite{moraes/2014}. However, here, a complete cosmological analysis of the model should explicit the form for which mass is varying with time. The objective of the next section is to provide a novel relation for $dm/dt$. Such a relation will reveal some important consequences in the realm of STM cosmology.

\section{New implications for the null condition application in Space-Time-Matter theory of gravity}\label{sec:novel}

In GR it is well known that the null condition applied to the line element ($ds=0$) describes the propagation of light in the presence of gravitation. It has been shown how to obtain the paths of particles from the null condition in STM\cite{wesson/2015,wesson/2013b,seahra/2001}. Indeed, such an application implies that all particles behave like photons in 5D whether or not they have rest mass in 4D., i.e., $dS^{2}=0$ replaces (4D) $ds^{2}\geq0$ as the basis for causality.

One might wonder the cosmological implications of the null condition application in STM. By taking $dS^{2}=0$ in Eq.(\ref{eqn:fes1}) yields

\begin{equation}\label{eqn:rel1}
\frac{dm}{dt}=\pm\frac{1}{\sigma^{n}}\sqrt{1-(\sigma v)^{2}},
\end{equation}
in which $v=dr/dt$, with $dr=\sqrt{dx^{2}+dy^{2}+dz^{2}}$, is considered the expansion velocity of the universe.

Eq.(\ref{eqn:rel1}) reveals a dependence of the mass variation on the expansion velocity of the universe, i.e., a more restrictive dependence than the age of the universe dependence originally found in STM and other variable-gravity theories. Such a dependence, as one can see, is mathematical, and not only conceptual or intuitive. 

\subsection{Implications for an accelerated expansion}\label{ss:pi}

Eq.(\ref{eqn:rel1}) can also bring up important features on the expanding universe dynamics if one knows the functional form for $m(t)$. 

It is worth remembering now what was mentioned in Section \ref{sec:stm}, i.e., that by expanding the STM geodesic equation, Wesson has found Eq.(\ref{eqn:stm1}) as the only component involving the fifth coordinate. The integration of (\ref{eqn:stm1}) results in $m=\kappa t+m_0$ with $\kappa$ and $m_0$ being constants. By substituting such a solution as well as (\ref{eqn:cs1}) in (\ref{eqn:rel1}), one is able to write the expansion velocity of the universe in STM as

\begin{equation}\label{eqn:pi1}
v=\frac{1}{\kappa t+m_0}\left[\frac{\eta}{(3n+1)t}\right]^{\eta}\sqrt{1-\left\{\kappa(\kappa t+m_0)^{n}\left[\frac{(3n+1)t}{\eta}\right]^{\eta n}\right\}^{2}}.
\end{equation}

At this stage it is worth mentioning that according to Type Ia supernovae observations, our universe is undergoing a phase of accelerated expansion\cite{riess/1998,perlmutter/1999}. Indeed, recent observation of anisotropies in the temperature of cosmic microwave background radiation support such a cosmological phenomenon\cite{hinshaw/2013}. 

The disability in matching the distance modulus of Type Ia supernovae predicted theoretically by GR with their observed values made Riess et al.\cite{riess/1998} and Perlmutter et al.\cite{perlmutter/1999} to assume an ``anti-gravitational'' constant term in the Einstein's FEs, known as cosmological constant. Such a term makes the universe to recently accelerate its own expansion. In fact, such an acceleration does not dominate the universe dynamics for the whole time. By compiling a list of independent measurements of the Hubble parameter, the transition - from a decelerated to an accelerated expansion phase of the universe - redshift was found to be $z_{\mathcal{T}}\sim0.74$\cite{farooq/2013}, that refers to an epoch for which the universe had approximately half of its present age. By using Eq.(\ref{eqn:pi1}), I am going to show that these cosmological features are predicted purely by the variation of mass - velocity of the universe expansion relation above.

Figure \ref{fig2} below depicts the time evolution of the acceleration of the universe expansion, i.e., $dv/dt$, with $v$ given by Eq.(\ref{eqn:pi1}).

\begin{figure}[]
\vspace{0.5cm}
\centering
\includegraphics[height=5.5cm,angle=00]{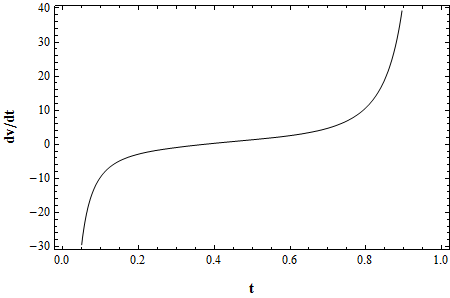}
\caption{Time evolution of the acceleration of the universe expansion from Eq.(\ref{eqn:pi1}) by taking $n=0.5$, $m_0=1$ and $\kappa=-1$.}
\label{fig2}
\end{figure}  
Note that the model predicts a period of negative acceleration prior to a transition epoch ($dv/dt\sim0$) followed by a positive acceleration phase. Until the universe reaches some epoch ($z_\mathcal{T}\sim0.74$), its density propitiated a matter-dominated dynamic scenario\footnote{In fact, for primordial epochs, the universe has also passed through inflation and radiation-dominated eras. An approach of the different phases of the universe through quintessence models has been recently presented\cite{moraes/2014c}.}. In other words, as the universe expanded, the gravitational force among its massive bodies, as clusters of galaxies, for instance, tended to act against the expansion, in this way, braking it and justifying $dv/dt<0$. From Figure \ref{fig2}, note that after a phase of transition, the model also predicts $dv/dt>0$, which in standard cosmology means a period of time in which the dynamics of the universe is dominated by a CC. 

\subsection{Implications for the variation of mass ceasing}\label{ss:mc}

Eq.(\ref{eqn:rel1}) also predicts that the variation of mass is, indeed, a transient phenomenom, since it ceases when the universe expansion velocity reaches $v=v_c=\sigma(t,m)^{-1}$. It can be noted, from Eqs.(\ref{eqn:cs1}) and (\ref{eqn:pi1}) along with the assumptions of the previous subsection, that $v\rightarrow v_c$ as $t\rightarrow 1$. Therefore, the variation of masses shall cease within a finite time, as a hitherto unpredicted STM feature.

\section{Discussion}\label{sec:dis}

The STM\cite{wesson/1983,wesson/1984, wesson/1985, wesson/1990, wesson/1992, wesson/2015} considers the GR as embedded in a higher dimensional surface, with the extra dimension being the mass. In fact, such an extra dimension is characterized by the rate at which masses vary with time. The effects of such a universe setup are negligible at non-cosmological scales, while in cosmological scales some important issues tend to rise. It is worth stressing that the variation of masses has been constantly approached both experimentally\cite{berengut/2013,kroupa/2001} and theoretically\cite{israelit/2007,israelit/2008,lima/2010,lima/2011,harko/2014}, however, the idea of considering it in the same footage as the 4D space-time dimensions was brought up by Wesson and since then, it has been improved by himself and collaborators.

As working in 5D, in this work, I have not considered a CC when developing the dynamical equations of the model. As once shown by Wesson, the CC in 4D is just an artefact of the reduction to 4D of the 5D KK models\cite{wesson/2001}. By dropping the CC, one, naturally, can get rid of the CC problem, which is the high discrepancy between the values of the quantum vacuum energy density obtained via cosmology\cite{hinshaw/2013} and via particle physics\cite{weinberg/1989}. As I shall stress below, even though no CC was considered in the universe composition, a cosmic acceleration was retrieved from the present model, just by allowing masses to vary.

In Section \ref{sec:fes}, I have derived some physical and cosmological parameters of the model from solution (\ref{eqn:cs1}). Note that Eq.(\ref{eqn:cs2}) although simple, is non-trivial. Even though $\sigma$ depends on both $t$ and $m$, $\rho$ remains restricted to the usual 4D manifold, since it does not present any dependence on the extra coordinate $m$. In other words, $\sigma(t,m)$ results in $\rho(t)$ only. The derived Hubble parameter also presents a valuable functional form according to (\ref{eqn:cs4}).

In Section \ref{sec:novel}, I have presented the relation (\ref{eqn:rel1}), which is a relevant STM novelty. Primarily, it tells us that rather than the age of the universe, the mass variation of the universe at cosmological scales depends directly on the velocity at which it expands. 

By expanding the STM geodesic equation, Wesson has obtained $d^{2}m/dt^{2}=0$. In other words, the rate at which mass varies with time should be a constant, which I named $\kappa$ in Subsection \ref{ss:pi}. In the particular case $\kappa=0$, one should recover, in its entirety, the conventional gravitational theory. In order to check this, let me assume $dm/dt=\kappa$ in (\ref{eqn:rel1}), so that one is able to write $v=\sqrt{1-(\kappa\sigma^{n})^{2}}/\sigma$. Then, if one takes $\kappa=0$ in $v$, one recovers the equation for the propagation of photons, i.e., the conventional theory is retrieved. In fact, such a retrieval supports the null condition application in STM, since it reveals the embedding property of GR in such a model, i.e., the standard gravity theory can be restored at any time by taking $dm/dt=0$, as originally required by Wesson.

The importance of Eq.(\ref{eqn:pi1}) became clear after plotting Fig.\ref{fig2}. The $\Lambda$CDM cosmological model describes different eras of the universe, such as radiation, matter and dark energy-dominated eras. Such a standard model universe if filled only by matter, could not predict the accelerated expansion. On the other hand, this work assumes the present universe is filled only by matter \footnote{The contribution from radiation can be neglected.} (remind I have assumed $\omega=0$ in (\ref{eqn:fes6}), which is the EoS of dust). In other words, the present work does not assume the existence of a CC, as expoused above, or any other entity added to the universe composition, with the purpose of describing the accelerated expansion. Nevertheless, from Fig.\ref{fig2} one can see the cosmic acceleration rising naturally just by characterizing $dm/dt\neq0$. 

It is worth noting that the universe expansion velocity will increase according to the model, independently of the sign of the deceleration parameter, because of re-collapse\cite{nojiri/2003}.\\
{\bf Acknowledgements}\

I would like to thank FAPESP for financial support. I would also like to thank the referee for bringing some references to my attention. 


\end{document}